# Slow-light-enhanced energy efficiency for the graphene microheater on silicon photonic crystal waveguides

Siqi Yan [1, 2], Xiaolong Zhu [2,3], Lars Hagedorn Frandsen [2], Sanshui Xiao [2,3], N. Asger Mortensen [2,3], Jianji Dong [1,*] and Yunhong Ding [2,*]

[1] Wuhan National Laboratory for Optoelectronics, Huazhong University of Science and Technology, 430074, Wuhan, China. [2] Department of Photonics Engineering and [3] Center for Nanostructured Graphene, Technical University of Denmark, DK-2800 Kongens Lyngby, Denmark

**ABSTRACT: Slow light has been widely utilized to obtain enhanced nonlinearities, enhanced spontaneous emissions, and increased phase shifts owing to its ability to promote light-matter interactions. By incorporating a graphene microheater on a slow-light silicon photonic crystal waveguide, we experimentally demonstrated an energy-efficient graphene microheater with a tuning efficiency of 1.07 nm/mW and power consumption per free spectral range of 3.99 mW. The rise and decay times (10% to 90%) were only 750 ns and 525 ns, which, to the best of our knowledge, are the fastest reported response times for**

**microheaters in silicon photonics. The corresponding record-low figure of merit of the device was 2.543 nW • s, which is one order of magnitude lower than results reported in previous studies. The influences of the graphene-photonic crystal waveguide interaction length and the shape of the graphene heater were also investigated, providing valuable guidelines for enhancing the graphene microheater tuning efficiency.**

After several decades of explosive growth, the information industry has become a major energy consumer owing to the high power consumption by data centers[1]. As one of the most promising candidates for satisfying the comprehensive information industry requirements of low energy consumption, speed, bandwidth, density and cost, integrated silicon photonics[2, 3, 4, 5] have made rapid progress in a wide range of functionalities such as modulators[6, 7], photodetectors[8, 9, 10] and optical switches[11]. This is largely enabled by silicon's low loss, low cost and high fabrication compatibility with complementary metal-oxide semiconductor (CMOS) technology[12, 13]. One of the most important properties of integrated devices in versatile and reconfigurable photonic networks is highly energy-efficient tunability with a fast response time. Owing to the relatively high thermo-optic coefficient of silicon ($\sim 1.8*10^{-4} K^{-1}$)[14], thermal tuning is often applied using a metallic microheater on a silicon waveguide in tunable silicon micro-ring resonators[15] (MRRs) or Mach-Zehnder interferometers[16] (MZIs). However, to avoid the light-absorption loss induced by the metal, a thick silicon dioxide ($SiO_2$) layer is typically introduced between the silicon waveguide and the metallic heater, inevitably impeding heat transport and dissipation owing to the low thermal conductivity of $SiO_2$ (1.44 W • $m^{-1}$ • $K^{-1}$)[17]. Several methods, such as the use of free-standing waveguide structures[18, 19] and different doping levels of the waveguide[20], have been proposed to simultaneously obtain lower power consumption and a faster response time.

However, free-standing waveguides may lack mechanical stability, and the tuning efficiency of different silicon waveguide doping is as low as 0.12 nm/mW[20].

The use of slow light in silicon photonic crystal waveguides (PhCWs) offers an approach for significantly improving the inherently weak light-matter interaction on nanometer-scale chips. By decreasing the group velocity of the transmitted light in periodic media[21, 22], slow light has been utilized in various applications such as sensors[23], amplifiers[24], and nonlinear optics[25]. Meanwhile, owing to many unique properties, such as a zero-band gap and tunable Fermi level[26, 27], high carrier mobility[28, 29] and ultra-broad absorption bandwidth[30], graphene has been widely merged with nanophotonic structures to enhance the light-matter interaction[31, 32, 33, 34]. In addition to these broadly studied applications, the use of graphene as a heating material[35, 36, 37] in close contact to the silicon waveguide can significantly improve the tuning efficiency due to graphene's low optical absorption rate[38]. Moreover, owing to the extremely high thermal conductivity of up to 5300 W·m$^{-1}$·K$^{-1}$, response times can be greatly reduced relative to devices with thick SiO$_2$ cladding between the metallic heater and the waveguides. However, the current performances of devices using graphene heaters are limited either by their relatively high power consumptions[36] or by their microsecond response times[37].

In this study, by incorporating graphene microheaters on silicon slow-light PhCWs, a graphene microheater with a high tuning efficiency and an ultra-fast response time is demonstrated. Based on the obtained data, the slow light effect can effectively enhance the thermal tuning efficiency. Furthermore, we also systemically investigate the influence of the graphene-PhCW interaction length and the shape of the graphene heaters on the tuning efficiency. The proposed slow-light-enhanced graphene microheaters show promising potential for use in integrated silicon building blocks such as tunable phase shifters and filters that demand low power consumption, a fast response time, and CMOS-compatible fabrication processes.

**Results**

**The design of the slow-light-enhanced graphene microheater**

A schematic of the slow-light-enhanced graphene microheater is shown in Fig. 1(a). A graphene monolayer is designed onto the core-region of the silicon PhCW. The graphene is contacted by two gold/titanium (Au/Ti) pads, exploiting the low contact resistance between Ti and graphene[39]. Ohmic heating is generated in the graphene via an applied voltage bias between the Au/Ti pads. The width of the graphene overlapping the photonic crystal line defect is designed to be narrower than the other part of the graphene to locally increase the Ohmic dissipation. This results in more effective heating, which will be discussed in the Discussion section. The inset in Fig. 2(a) illustrates the temperature distribution of the cross section of the proposed microheater when the graphene is powered up, indicating a tight localization of the thermal fields in the silicon membrane.

The use of slow-light can enhance the tuning efficiency owing to the large group index that can be obtained in the PhCW, which increases the effective interaction length between the heater and the waveguides[22], thus increasing the corresponding phase shift. The phase shift $\Delta\varphi$ induced by the heating can be expressed as

$$\Delta\varphi = \Delta T * a * \left(\frac{\omega}{c}\right) * n_g * L \qquad (1)$$

In Eq. (1), $\Delta T$ is the temperature change of the photonic crystal membrane, a is the thermo-optic coefficient of silicon, $n_g$ is the group index of the PhCW and $L$ is the graphene-PhCW overlap length. ω is the angular frequency of the light at the wavelength of 1550 nm. As we can see, increasing the group index decreases the temperature increase required to induce a $2\pi$ phase shift, leading to lower power consumption because the required power is proportional to the required temperature increase. A longer PhCW can also reduce the power consumption but will increase the insertion loss and the size of the device. The very high group index (larger than ~40) of the PhCW can only be achieved near the fundamental mode cut-off in a very narrow bandwidth where large insertion losses[21] are typical because slow-light promotes further damping[40]. Therefore, the band structure of the PhCW employed here should be carefully optimized to reach a relatively high group index, large bandwidth and low loss in the wavelength regime of interest. These requirements can be satisfied simultaneously using semi-slow-light PhCWs[41, 42].

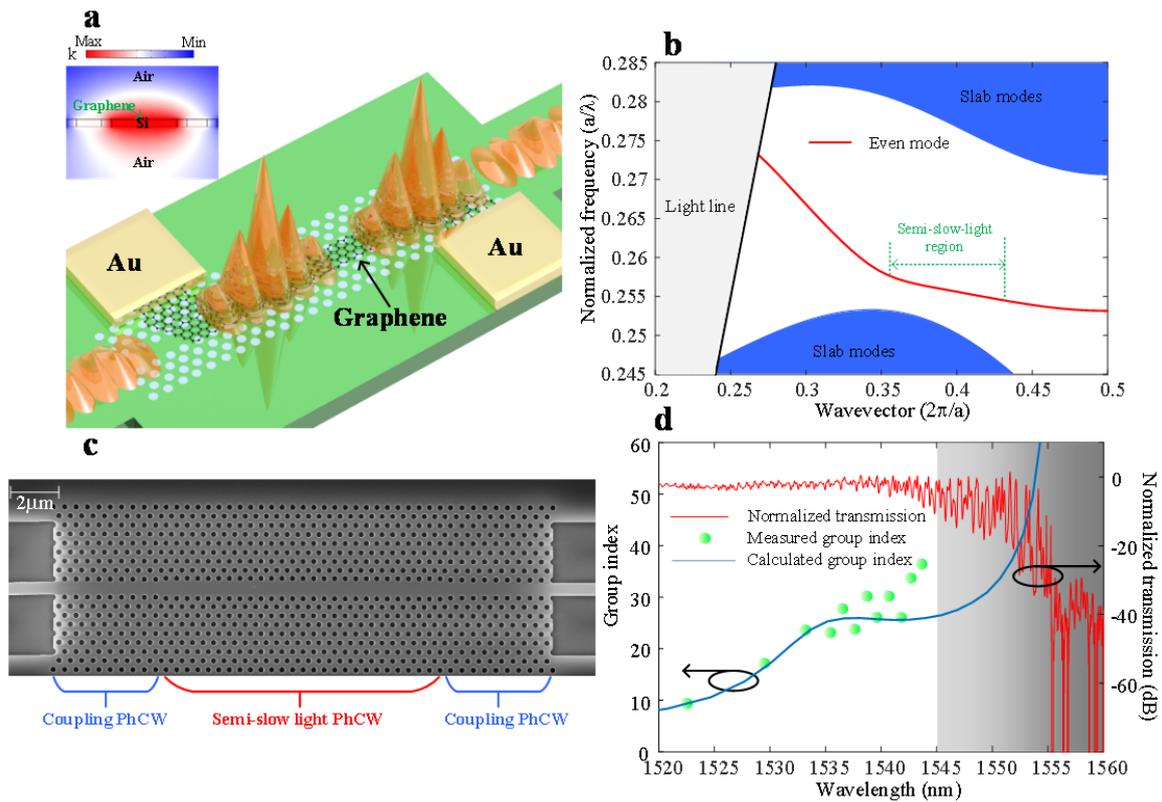

**Figure 1.** (a) Schematic of the slow-light-enhanced graphene heater. The inset shows the temperature distribution in the cross-section of the photonic crystal waveguide with a graphene microheater. (b) Band structure of the photonic crystal waveguide. The even guiding mode (red curve) consists of a semi-slow-light region (green dashed line). (c) SEM image of the fabricated photonic crystal waveguide. (d) Calculated group index (blue curve), measured group index (green dots) and the transmission spectrum (red curve) of the photonic crystal waveguide.

The dispersion relationship of the PhCW is analyzed using the plane-wave-expansion method[43]. To obtain a high group index with a large bandwidth, the positions of the first and second rows of holes adjacent to the line-defect are slightly tuned in a W1-type PhCW. After optimization, the dispersion relationship of the PhCW is presented as the red line in Fig. 1(b). We only focus on the even symmetry PhCW mode. In our final design, the lattice constant is set as 390 nm, and the diameter of the holes is 193 nm, respectively. The position of the first row of holes adjacent to the PhCW core is moved 41 nm outward from the original position, and the second row is moved 10 nm outward. Figure 1(c) shows a scanning electron microscope (SEM) image of the fabricated silicon PhCW, where coupling regions are introduced between the strip waveguide and slow-light PhCW to reduce the coupling loss[44]. Figure 1(d) displays the calculated (blue line) and measured (green dot) group indices as well as the measured transmission spectrum (red line). As shown by the green curve, the fabricated silicon PhCW has a semi-slow-light region at approximately 1540 nm with a group index of approximately 25; a larger group index can be obtained at longer wavelengths at the cost of larger insertion losses and pronounced fluctuations with wavelength, as shown in the gray area in Fig. 1(d). The slight deviation from the theoretical calculation is attributed to the deviations in the fabrication procedure as well as the presence of the thin aluminum oxide ($Al_2O_3$) layer deposited on the silicon that leads to the tiny shift of the group index curve from the ideal curve to the curve with the longer wavelength range. The trade-off for the slow-light enhancement is a reduced spectral operation bandwidth, which is optimized to approximately 10 nm in our design.

**Characterization of the slow-light-enhanced graphene microheater**

To characterize the performance of the graphene microheater, we fabricated a tunable MZI filter on a silicon-on-insulator (SOI) wafer with a 250 nm thick top silicon layer and a 2 μm SiO$_2$ buried oxide (BOX) layer (Supporting Information Section 2), as shown in Figs. 2(a) and 2(b). The TE-polarized light is coupled from a fiber to the waveguide using a photonic crystal grating coupler[45]. A multi-mode interferometer (MMI) divides the input light equally into the two arms of the MZI, both consisting of silicon strip waveguides with dimensions of 450 nm*250 nm and 20 μm long PhCWs. One of the strip waveguides is designed to be 145.7 μm longer than the other in order to induce an appropriate optical path difference. Moreover, to balance the loss induced by the presence of graphene, the PhCW in both arms have equal lengths of graphene on top.

The transmission spectrum (blue line) of the fabricated MZI without the graphene heater is depicted in Fig. 2(c). In our analyses, the measured optical power output is normalized to the reference strip waveguide to exclude the coupling loss of the grating couplers. According to Fig. 1(d), the interference dip at 1533.71 nm lies in the slow-light region. The measured FSR is approximately 4 nm. For comparison, the transmission spectrum of the device incorporating the graphene heater is measured as well, which is shown in the red curve in Fig. 2(c). An excess loss of 5 dB is induced, and the extinction ratio is degraded to approximately 8 dB in the MZI with graphene, which is attributed to metallic contamination during the wet-transfer and lift-off process[46]. Such degradations can be optimized by improving the wet-transfer process by using a modified Radio Corporation of America (RCA) clean process[47].

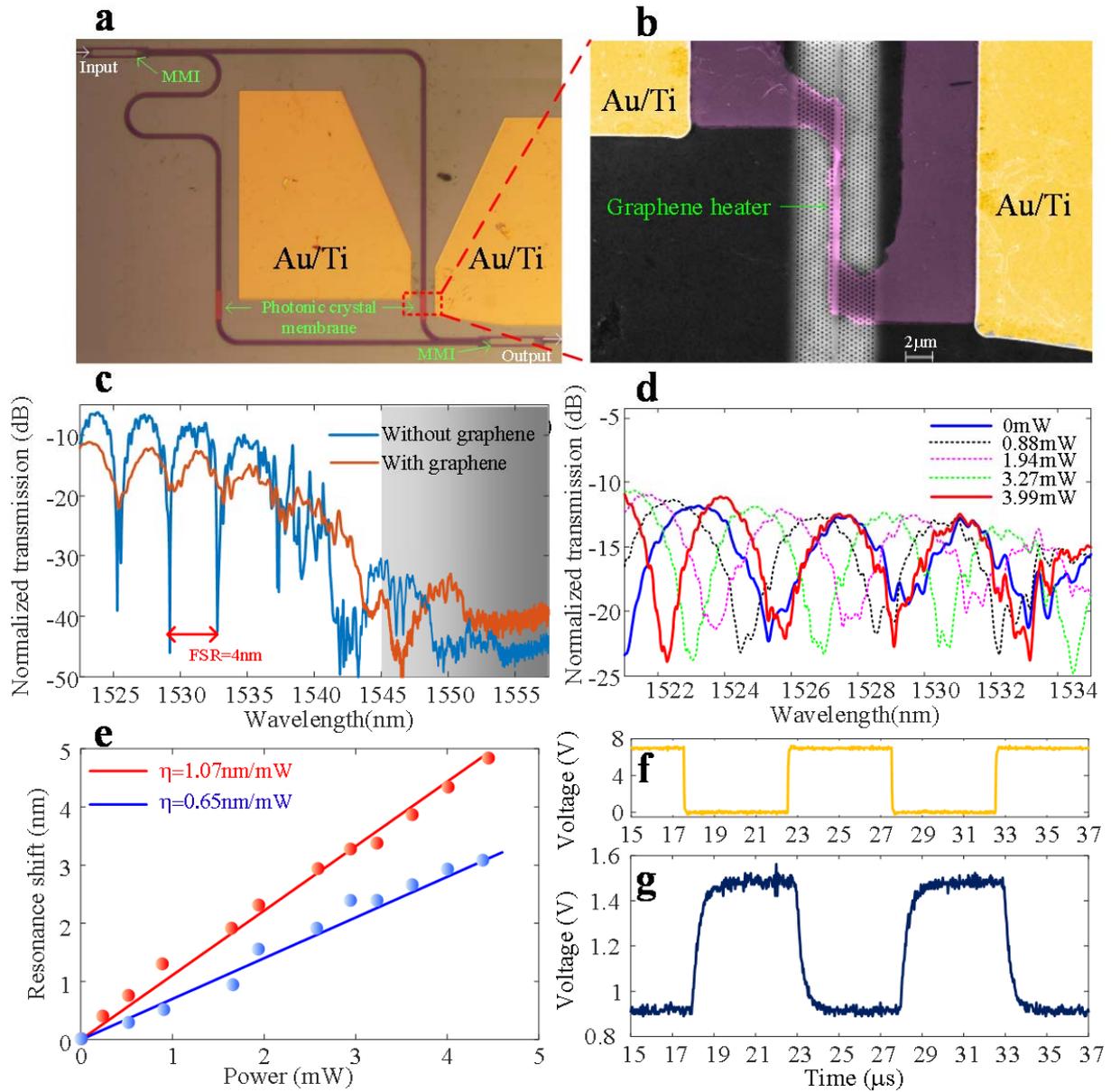

**Figure 2.** (a) Microscope image of the entire MZI device. (b) False-color SEM image of the slow-light-enhanced graphene heater. (c) Measured and normalized transmission spectra for the MZI with (red line) and without (blue line) graphene. In the gray area, the light lies in the bandgap of the PhCW. (d) Static response of the heating power. (e) Measured resonance shifts for the interference dips at 1525.12 nm (blue) and 1533.71 nm (red) as functions of the applied heating power. (f) Driving electrical signal and corresponding temporal response signal (g).

Next, an external voltage is applied to the graphene heater, and the spectral responses are measured for different applied powers; the results are presented in Fig. 2(d). The shift of the interference dip at approximately 1533.71 nm reaches one FSR (from the solid blue line to the solid red line) with a tuning power of only 3.99 mW. Meanwhile, there is a larger shift of the interference dip at 1533.71 nm than of the dip at 1525.12 nm, indicating a tuning efficiency enhancement induced by the slow-light effect. Figure 2(e) plots the shifts of the interference dips at 1533.71 nm (red line) and 1525.12 nm (blue line) as functions of the tuning power, and tuning efficiencies ($\eta$) of 1.07 nm/mW and 0.65 nm/mW are achieved, respectively.

The response time of the graphene heater is further characterized by driving the graphene heater with a square waveform electrical signal while the wavelength of the input signal is fixed at 1531.8 nm. The frequency of the driving signal is set to 100 kHz, and the peak-to-peak voltage ($V_{pp}$) is set to 7 V, as shown in Fig. 2(f). The $V_{pp}$ of 7 V corresponds to an applied heating power of 3.27 mW. The modulated light is received by a photodetector and recorded by an oscilloscope; the results are shown in Fig. 2(g). The 10% to 90% rising and decaying times are measured to be 750 ns and 525 ns, respectively. To the best of our knowledge, this is the fastest response time reported for a microheater in silicon photonics. The fast response time is attributed to the extremely high thermal conductivity of graphene and the membrane structure without underlying thick buried oxide layer. The figure of merit (FOM), i.e., the product of the power consumption per FSR and the average response time[48], can be used for a comprehensive evaluation of the microheater performance. Our slow-light-enhanced graphene heater has a FOM as low as 2.543 nW·s, which is one-order of magnitude better than previous demonstrations[48].

**The influence of the graphene-PhCW interaction length.**

To investigate the impact of the graphene-PhCW interaction length on the energy efficiency of the heater, we fabricated and measured the MZI with graphene-PhCW overlap lengths of ~5 μm and ~3 μm, shown as the insets in Figs. 3(c) and 3(d). The measured transmission spectra at different heating powers for graphene-PhCW overlap lengths of ~5 μm and ~3 μm are shown in Figs. 3(a) and 3(b), respectively, and the corresponding shifts of the transmission spectrum dips as functions of the heating power are depicted in Figs. 3(c) and 3(d). The tuning efficiency decreases while the power consumption per FSR increases for the shorter graphene-PhCW overlap length, which is in accordance with Eq. (1).

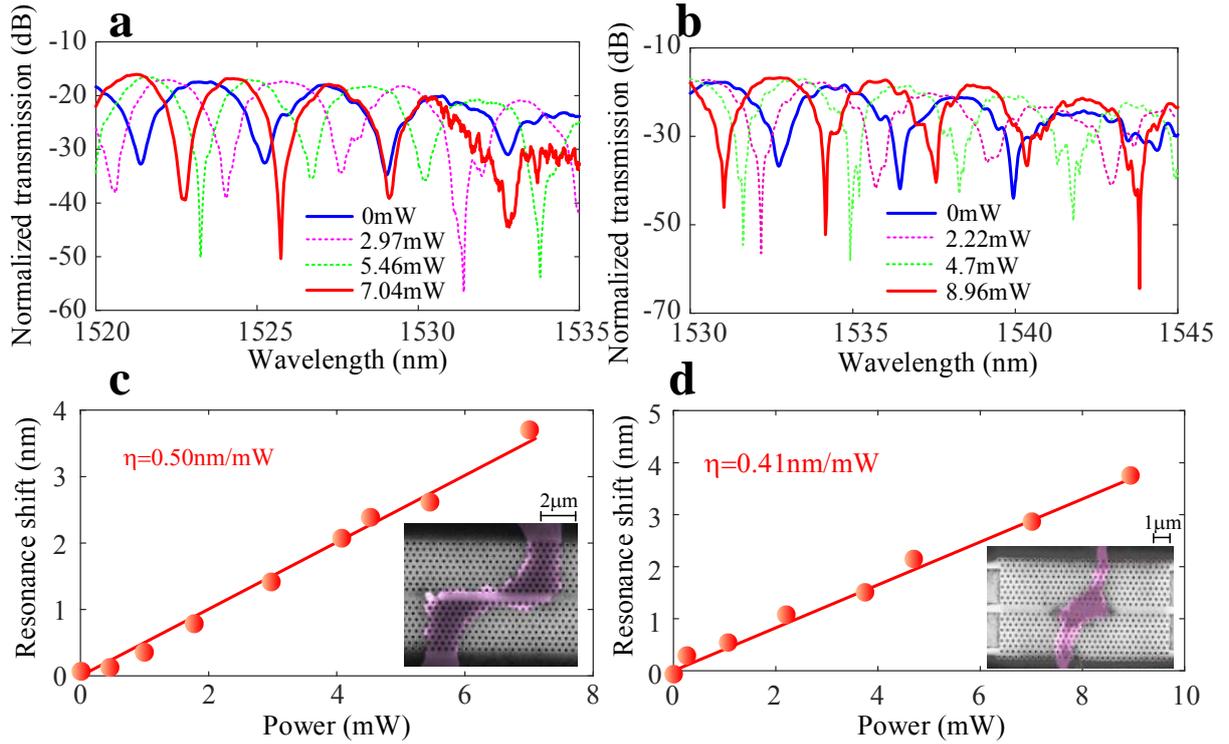

**Figure 3.** Transmission spectra at different heating powers for devices with graphene-PhCW interaction lengths of (a) ~5 μm and (b) ~3 μm. (c, d) Corresponding shifts of the transmission spectra as a function of the heating power, where the insets are false-color SEM images of the devices with ~5 μm and ~3 μm graphene-PhCW interaction lengths.

**Discussion**

The shape of the graphene heater can have a significant influence on the tuning efficiency. As shown in Fig. 2(b), in the previous design, the graphene coverage on the photonic crystal is designed to be 'Z-shaped' to boost the tuning efficiency. For comparison, a 'straight-shaped' graphene heater fully covering the PhCW membrane is fabricated, as shown in Fig. 4(a). Figure 4(b) shows the voltage-current relationships of both the straight-shaped (red line) and Z-shaped (blue line) graphene layers. The total resistance of the straight-shaped heater is 1.42 kΩ, which is more than ten times lower than that of the Z-shaped heater. According to the data in Fig. 4(c), the heating efficiency is only 0.14 nm/mW for the straight-shaped heater, much lower than that for the Z-shaped heater. This is because in the Z-shaped case, heat is predominantly generated in the region overlapping the photonic crystal membrane where the light is mainly confined. Thus, the heating-induced modulation of the waveguide can be much more efficient. In contrast, in the straight-shaped case, heat is generated uniformly throughout the entire graphene layer, and only a very small part of the heater effectively overlaps the optical mode; hence, the tuning efficiency is relatively modest. We also test the dynamic response of the device with the straight-shaped heater using the same square waveform electrical signal shown in Fig. 4(d) and the modulated signal shown in Fig. 4(e). Here, we find the rise and decay times to be 850 ns and 875 ns, respectively. The slower rising time is due to the low heating efficiency, and the slower decay time may be because more overall heat is generated with the same external voltage, requiring more time for dissipation of the heat through the silicon membrane.

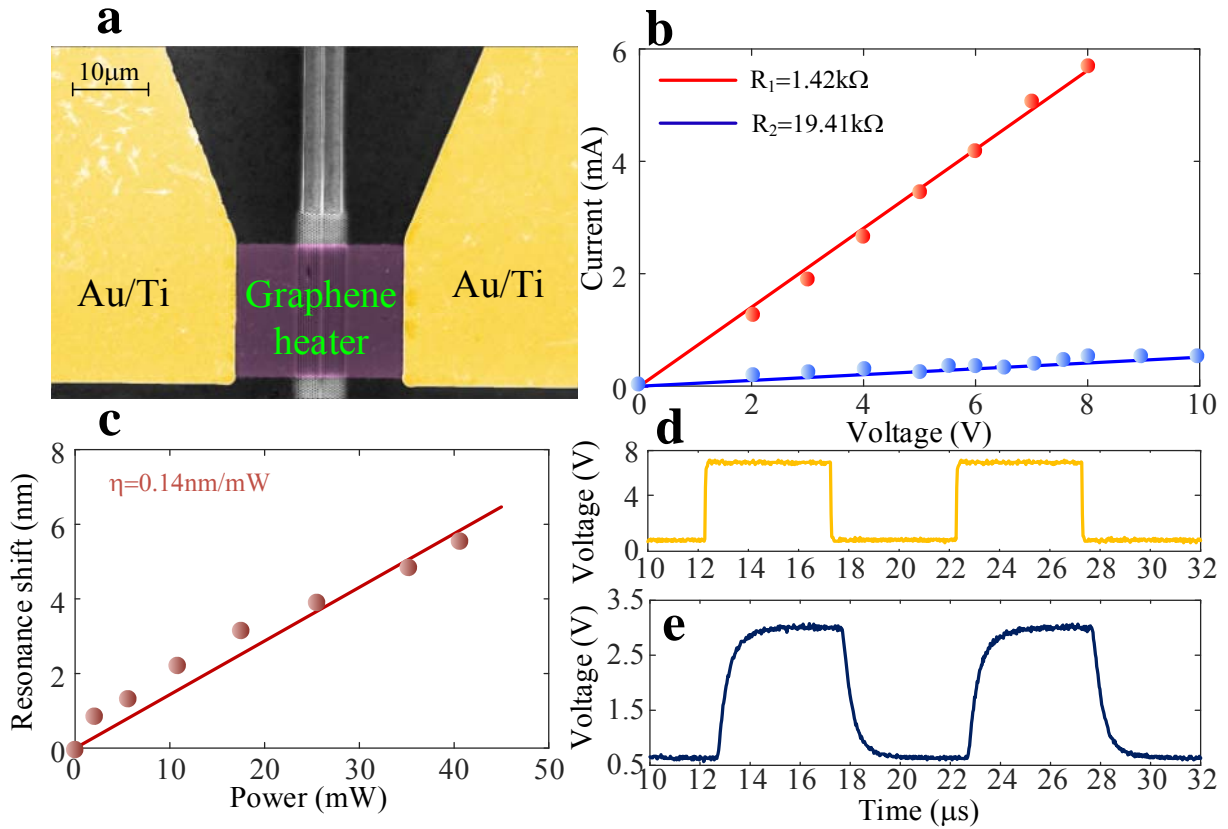

**Figure 4.** (a) False-color SEM image of the straight-shaped graphene heater. (b) Voltage-current relationship for the (blue line) Z-shaped and (red line) straight-shaped heaters. (c) Shift of the transmission spectrum as a function of the applied heating power for the straight-shaped heater. (d) Driving electrical signal and (e) corresponding modulated output signal of the device.

In conclusion, we have experimentally demonstrated slow-light-enhanced energy-efficient graphene microheaters with ultrafast response times. We have comprehensively studied the influences of the graphene-PhCW interaction length and the shape of the graphene heater. Owing to the slow-light effect, the heating efficiency is as high as 1.04 nm/mW, and the power consumption is as low as 3.99 mW. Furthermore, the 10% to 90% rising and decaying times are measured to be only 750 ns and 525 ns, which are the fastest ever reported for microheaters in silicon photonics. To the best of our knowledge, the figure of merit of the proposed device is lower than all previously demonstrated filters based on microheaters. The relatively high insertion loss could be greatly improved by optimizing the fabrication process to eliminate the metallic contamination during the wet-transfer process[49, 50]. The CMOS-compatible fabrication process of the proposed device enables its wafer-level integration with other nanophotonic devices. The slow-light-enhanced energy-efficient graphene microheater has demonstrated its distinctive advantages in simultaneously achieving both a low power consumption and a fast response time, showing great potential for its use in configurable photonic integrated circuits.

**Method:**

The proposed tunable MZI filter was fabricated on an SOI wafer with a 250 nm thick silicon layer on top of a 2 μm $SiO_2$ buried layer. E-beam lithography (EBL) and inductively coupled plasma (ICP) etching were used to fabricate the grating couplers, strip waveguides, MMI and the photonic crystal waveguides. Standard ultraviolet (UV) lithography was performed to define the wet etch regions used to undercut the photonic crystal waveguides with AZ5124E acting as the mask. Buffered hydrofluoric acid (BHF) was used to etch the $SiO_2$ buried layer below the photonic crystal waveguides. After the membranization, an 11 nm $Al_2O_3$ layer was deposited on the device by atomic-layer deposition (ALD). A graphene sheet grown by CVD was wet-transferred onto the silicon device. In the wet-transfer process, AZ resist was first spin-coated onto the graphene covered copper foil and dried at 100°C for 1 min. Next, an AZ/graphene membrane was obtained by etching away the copper foil in an $Fe(NO_3)_3/H_2O$ solution and then transferring onto the silicon chip. Then, the AZ resist was dissolved in acetone, simultaneously cleaning the graphene surface. The graphene shape was defined using standard UV lithography and $O_2$ plasma etching. Finally, Au/Ti contacts were fabricated on the graphene by standard UV lithography, followed by metal deposition and a lift-off process.

**Corresponding Author**

*Email: jjdong@mail.hust.edu.cn and yudin@fotonik.dtu.dk

**Author Contributions**

S. Y. and Y. D. proposed the slow light enhanced tunable silicon MZI with a graphene heater. S. Y., L. H. F. and S. X. performed numerical simulations. S. Y., Y. D. and X. Z. fabricated the

graphene−silicon hybrid tunable MZI device. S. Y. and Y. D. performed the measurements. S. Y., Y. D., S. X. and J. D. discussed and analyzed the measured data. S. Y. wrote the first draft of the manuscript. All the authors discussed the results and contributed to the writing of the manuscript.

**Funding Sources**

This work is supported by the Danish Council for Independent Research (DFF-1337−00152 and DFF-1335−00771). The Center for Nanostructured Graphene is sponsored by the Danish National Research Foundation, Project DNRF103. S. Yan is sponsored by the China Scholarship Council (CSC) for supporting his work in Denmark.

**Notes**

The authors declare no competing financial interest.